\documentclass{PoS}

\title{Fast variability from X-ray binaries}

\ShortTitle{Fast variability from X-ray binaries}

\author{\speaker{Tomaso M. Belloni}\\
        INAF - Osservatorio Astronomico di Brera\\
        Via E. Bianchi 46, I-23807 Merate, Italy\\
        E-mail: \email{tomaso.belloni@brera.inaf.it}}

\abstract{The X-ray emission from accreting black-holes and neutron stars features strong variability on sub-second time scales, with very complex and broad phenomenology. From high-frequency quasi-periodic oscillations to rapidly changing X-ray burst oscillations to millisecond pulsations, these are weak signals immersed in strong noise and their study is pushing instrument capabilities to their limit. The scientific significance of fast time variability studies are both astronomical (properties of accretion flows, nature and evolution of sources) and physical (effects of General Relativity, equation of state of degenerate matter). I first review the main observational properties, then discuss the future prospects and observational needs.
}

\FullConference{High Time Resolution Astrophysics IV - The Era of Extremely Large Telescopes - HTRA-IV,\\
		May 5-7, 2010\\
		Agios Nikolaos, Crete, Greece}

\begin{document}

\section{The Promise of X-ray Binaries}

\begin{figure} 
\begin{center}
\includegraphics[width=.8\textwidth]{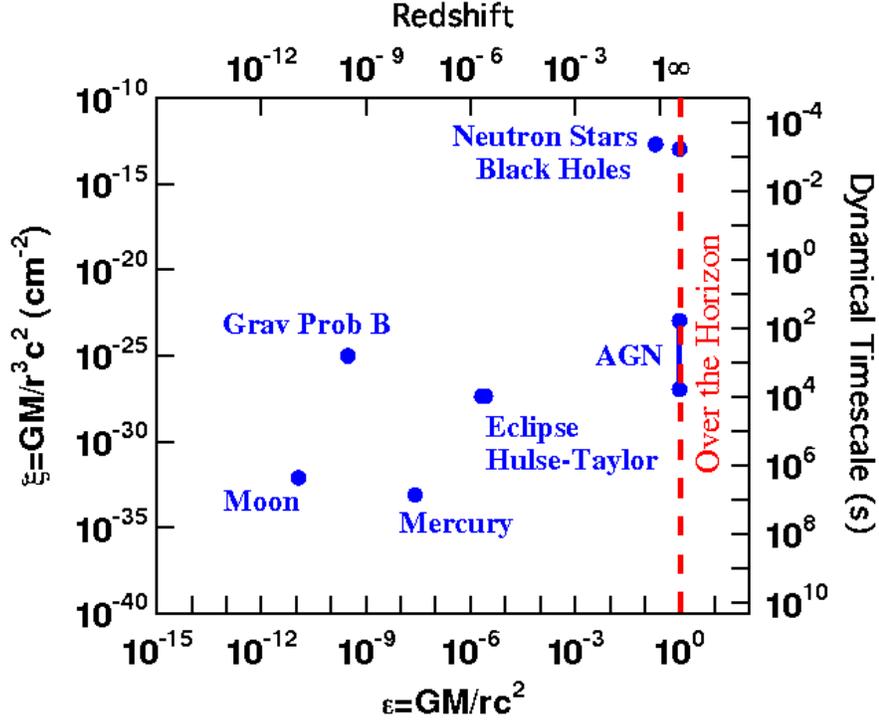} 
\caption{Tests of General Relativity placed on an appropriate
  parameter space.  The $x$-axis measures the potential
$\epsilon\equiv GM/rc^2$ and the $y$-axis measures the spacetime
curvature $\xi\equiv GM/r^3c^2$ of the gravitational field at a radius
$r$ away from a central object of mass $M$.
 The long-dashed line represents the event horizon
  of Schwarzschild black holes (from \cite{psaltis2008})} 
\label{fig:psaltis} 
\end{center}
\end{figure}

X-ray binaries are systems made of a compact object and a non-collapsed star. When dealing with high-energy emission however, it is more correct to say that they are made of a compact object and matter orbiting around it and falling onto/into it. They therefore 
constitute an ideal laboratory for studying two separate phenomena: accretion onto the compact object and General relativity in the strong field regime. Matter orbiting a few kilometers from a neutron star or a black hole is part of a complex accretion flow and is located in the deepest part of the gravitational well. The best way to study a gravitational field is to make use of a test particle and here we have many such test particles right where we would like to have them.  While active galactic nuclei also provide a comparable gravitational potential, the field curvature is much smaller than in galactic systems (see Fig. \ref{fig:psaltis}, from \cite{psaltis2008}). Therefore, X-ray binaries appear as the best space laboratories.

The main problem with these approaches is that they are entangled: we want to understand the properties of accretion close to a compact object making use of the fact that they are in a strong gravitational field, while at the same time we want to study the same gravitational field making use of the accreting matter. Accretion onto a compact object is a very complex and messy phenomenon: it provides not a test particle, but a stream of magnetized plasma that moves towards the compact object forming a very complex object. This intrinsic difficulty has not yet been overcome, but the field remains very promising.

In addition to probing General Relativity, neutron-star binaries are systems where we can detect strong high-energy emission from the surface of the neutron star and its surroundings. This provides a potential instrument to put limits on the equation of state of neutron matter, in particular as it offers the possibility of measuring mass and radius of the central object.

Both these goals have still to be reached. They are approached from two separate directions: the spectral distribution of the X-ray emission and the fast time variability. As continuum energy spectra of neutron-star Low-Mass X-Ray Binaries (LMXB) are very complex and do not offer a simple and unique interpretation, the attention has been concentrated on the shape of the iron emission fluorescence line due to reflection of X-ray radiation off the accretion disk in the system. Since the expected line is narrow, the effects of relativity (such as gravitational redshift, transverse doppler shift and beaming) on the orbiting gas, smeared by the complex motion in the accretion disk, can be studied.
The alternative approach is through fast timing. Timing signals are the most direct way of studying the motion of matter around a compact object and there are plenty of these signals observed. However, in this case the theoretical modeling is still not able to interpret the complex phenomenology observed in this variability. 
Below, I briefly outline the current standpoint and the its future prospects.

\section{Fast variability}

\subsection{Time scales}

When dealing with time variability, it is important to consider the possible characteristic time scales that we can expect from the system. In addition to the spin period of the neutron star in the case of a pulsar, an obvious frequency that could be (but is not always) observed, we have all the time scales associated to the accretion flow (see e.g. \cite{belloni1997a,nowak1999a,nowak1999b}). Important quantities are e.g. the time scales for radial light-crossing, radial sound-crossing, free-fall, viscous and thermal diffusion. The identification of these time scales in the data would provide important insights on the process of accretion (see e.g. \cite{belloni1997b}).
In addition, there are the time scales associated to General Relativity: a particle orbiting around a compact object with angular momentum, in addition to its Keplerian orbital period, is also subject to nodal and frame-dragging precessions. The identification of one or more of these characteristic time scales in the emitted radiation and its firm identification is the most promising path to the discovery of relativistic effects. However, the situation is very complex and it is difficult to disentangle the various components.

\subsection{Pulsations}

\begin{figure} 
\begin{center}
\includegraphics[width=.8\textwidth]{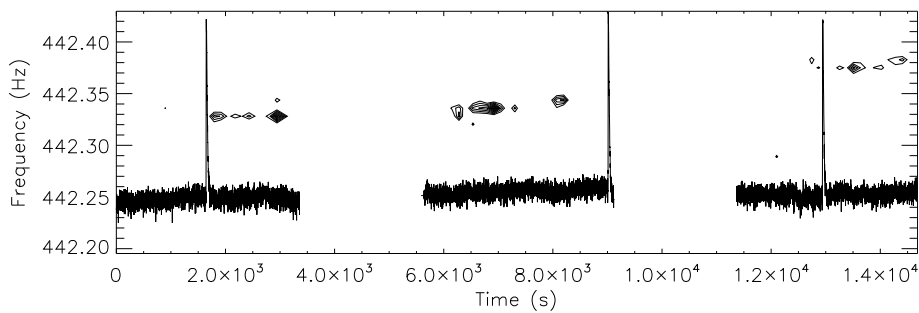} 
\caption{
Spectrogram of a RXTE observation of the intermittent pulsar SAX~J1748.9--2021. The contours show the pulsation, whose frequency drifts due to orbital modulation.  Superimposed is the light curve, which shows the presence of X-ray bursts (from \cite{diegoqpo}).
  } 
\label{fig:altamirano} 
\end{center}
\end{figure}

Although X-ray pulsars are known since the dawn of X-ray astronomy, the ``normal'' systems of this type have magnetic fields so high that they do not allow the formation of an inner accretion flow and start channeling matter onto the neutron star at a large distance. However, since 1999 we have discovered a few systems with low-mass companion showing millisecond pulsations (see \cite{michiel2006}). The discovery of these pulsations has been a goal for decades, when people looked into the emission of bright sources, where no detection was obtained. As it happens, pulsations are detected from faint transients and are associated to very compact binaries. More recently, intermittent pulsations have been detected from a few systems, including the bright transient Aql X-1 (\cite{casellaqpo,diegoqpo}). In Aql X-1, the pulsation appeared for a mere 150 seconds over a total exposure time of 1.3$\times 10^6$s. Another example where more pulsation intervals were detected is shown in Fig. \ref{fig:altamirano}. It is not yet clear what leads to these intermittency and what prevents the detection of pulsations in other bright systems (for which we know the pulse period nevertheless, see below).

\subsection{Oscillations during thermonuclear X-ray bursts}

Neutron-star low-mass X-ray binaries (LMXB) at low accretion rate often show thermonuclear X-ray bursts. They are powerful strong surges in X-ray emission caused by the ignition of a thermonuclear reaction in the material accreted onto the surface of the neutron star (see e.g. \cite{duncan}). With the RossiXTE satellite, transient highly coherent (although drifting in frequency) oscillations were observed during some bursts in a number of sources and it was soon realized that the oscillation frequency was the same for all bursts of each source, suggesting that what was seen was the pulse frequency. This was later confirmed with the detection of pulsations and burst oscillations at the same frequency in the same source. This not only allows to enlarge the sample of sources for which we know the pulse period, but offers a very interesting window onto processes that take place on the very surface of the neutron star (see \cite{strohbild}). These signals, in the range $\sim$200-800 Hz, can be extremely weak and difficult to analyze. Moreover, the frequency evolution in time can be so fast that our observations are limited by the uncertainty principle.

\begin{figure} 
\begin{center}
\includegraphics[width=.8\textwidth]{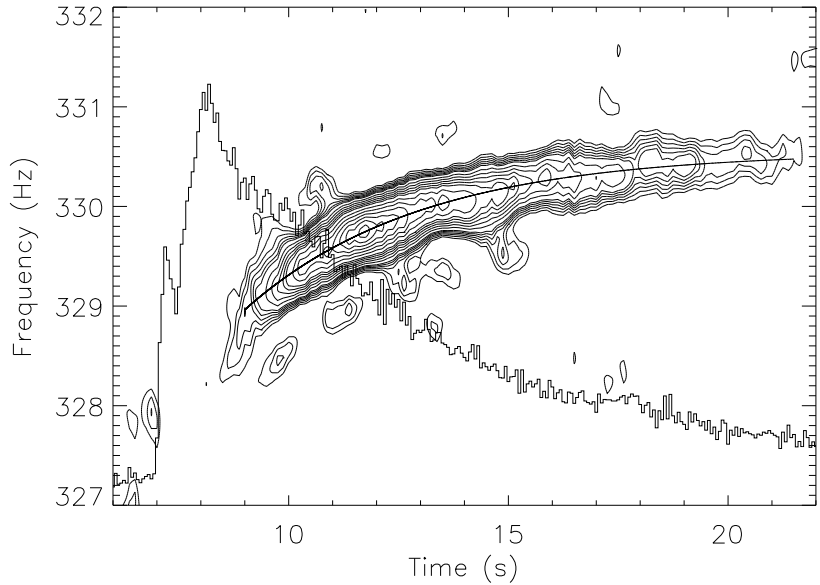} 
\caption{
Spectrogram for an X-ray burst of 4U~702---429, plotted over the RXTE/PCA light curve
(solid histogram). The best fit frequency evolution is shown as a thick solid line (from \cite{strohbild}).
} 
\label{fig:burst} 
\end{center}
\end{figure}

\subsection{Noise variability}

\begin{figure} 
\begin{center}
\includegraphics[width=.9\textwidth]{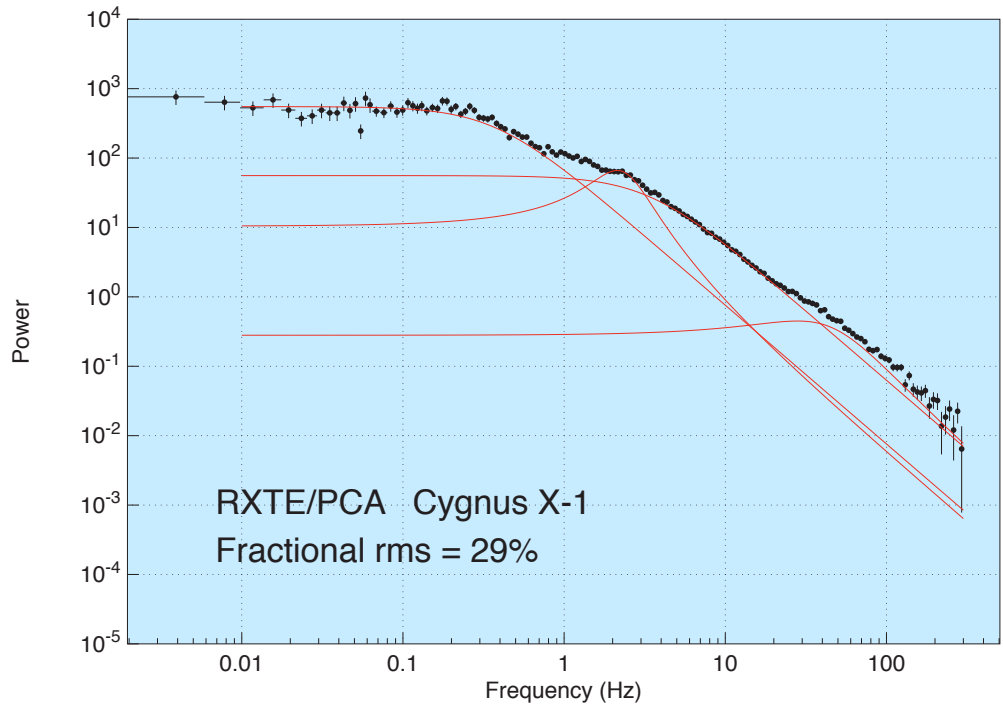} 
\caption{
Typical power-density spectrum of Cyg X-1 in its low/hard state. Three broad Lorentzian components (red) are overplotted.
} 
\label{fig:cigno} 
\end{center}
\end{figure}

\begin{figure} 
\begin{center}
\begin{tabular}{cc}
\includegraphics[width=.5\textwidth]{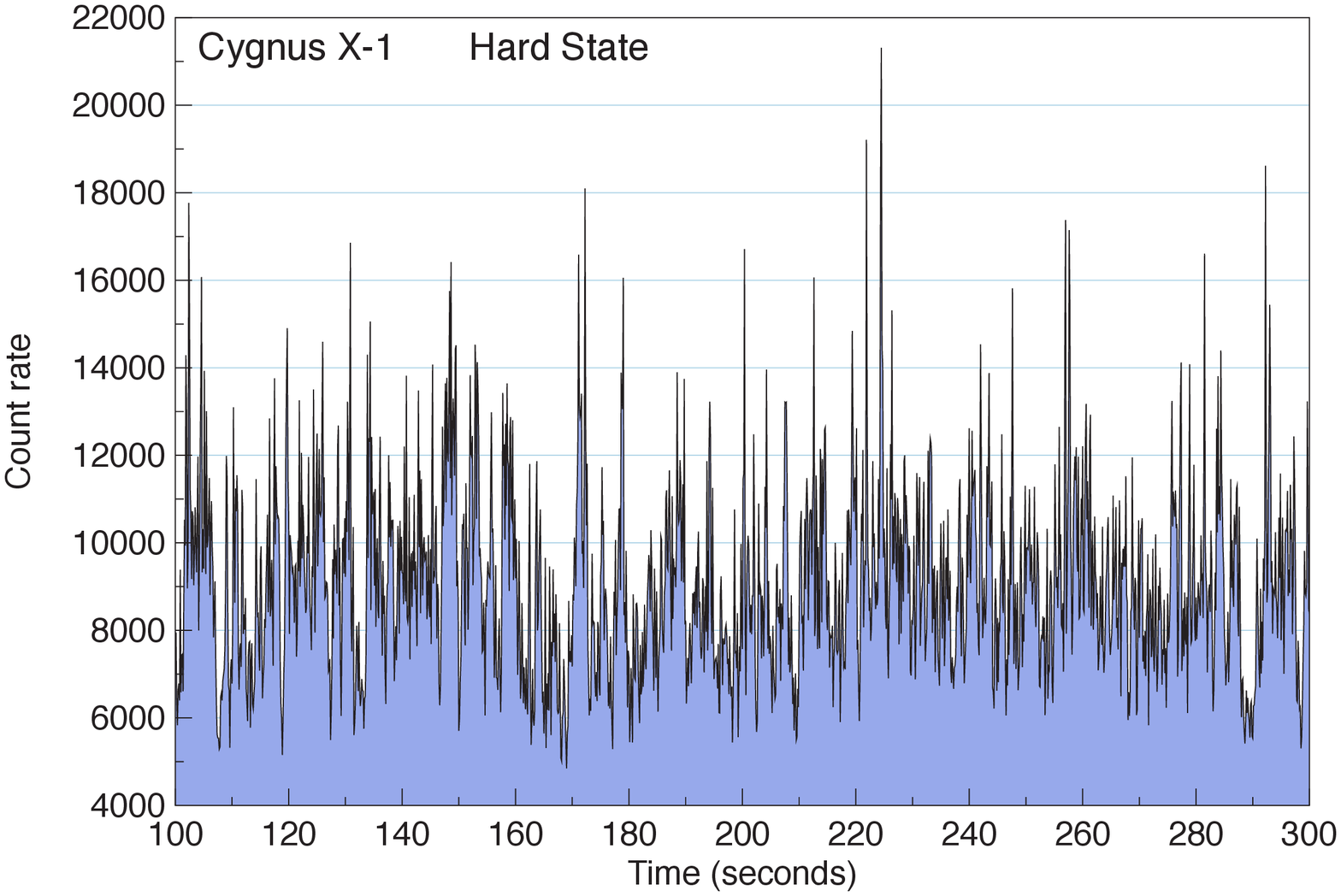} & \includegraphics[width=.5\textwidth]{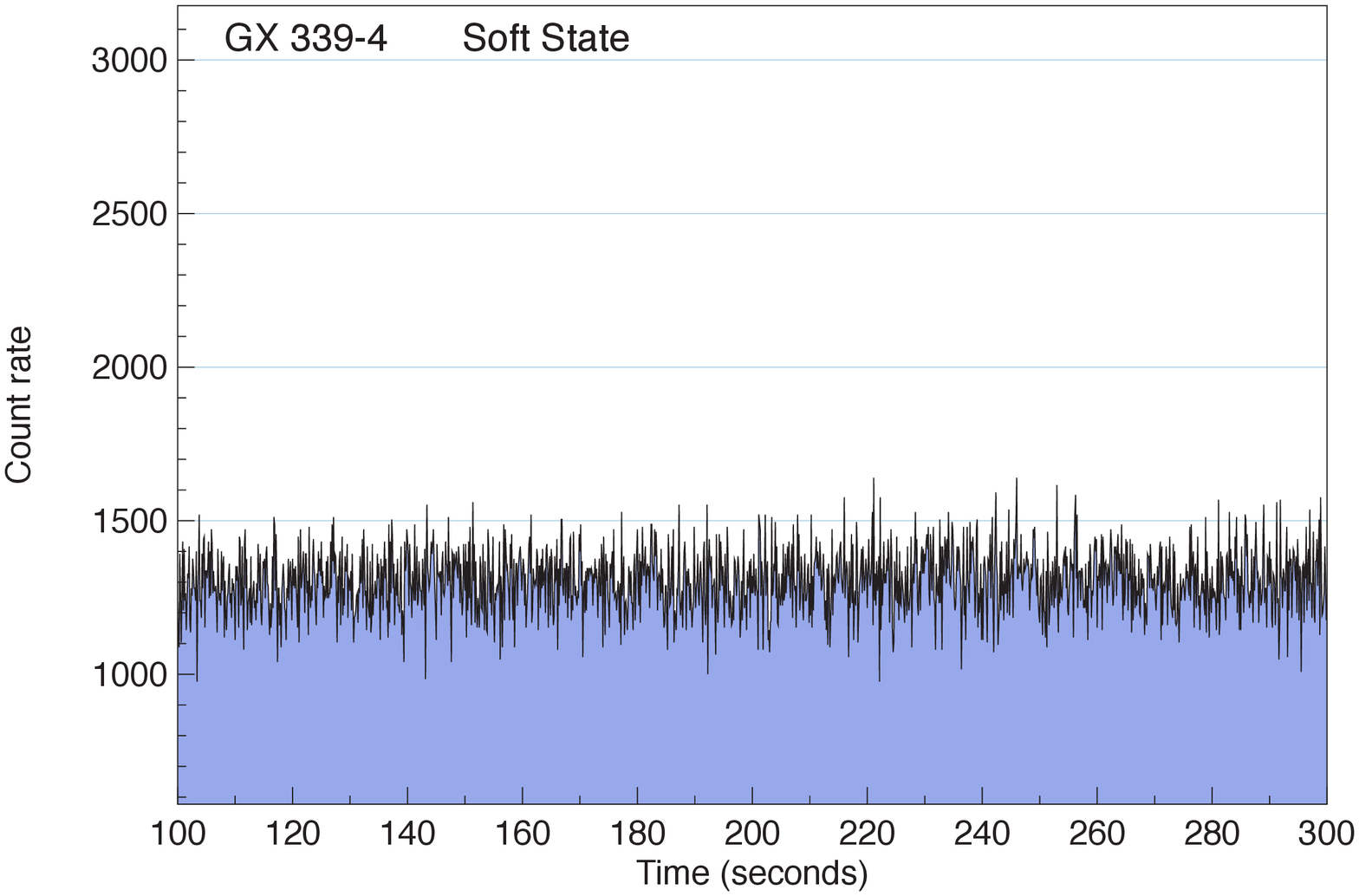}\\
\end{tabular}
\caption{
Left panel: a RXTE/PCA observation of Cygnus X-1 in the low-hard state; Right panel: a RXTE/PCA observation of GX 339-4 in the high-soft state. The tow panels have the same dynamical range in count rate. The difference in total variability is evident.
} 
\label{fig:hardsoft} 
\end{center}
\end{figure}

When in their low-luminosity states, both black-hole and neutron-star X-ray binaries, in addition to a rather hard energy spectrum, are characterized by very strong aperiodic variability, with an integrated fractional rms that can reach 40-50\% (see \cite{belloni2010,gilfanov2010}). In the Fourier domain, this variability can be interpreted as the sum of a small number of Lorentzian-shaped components (\cite{olive1998,pottschmidt,nowak2000,bpk,belloni2010}, see Fig. \ref{fig:cigno}). These components move in frequency hand-in-hand, positively correlated with the source flux (\cite{pottschmidt,bpk}). At higher fluxes, their fractional rms decreases and Quasi-Periodic Oscillations appear (see below). This strong variability is associated to very clear signatures in the phase-lag spectrum, providing a complex but important set of constraints to theoretical models, whether they involve themal/hybrid Comptonization or contributions from a jet component (\cite{nowak1999a,gilfanov2010}).
In other states, where the emission is dominated by a thermal optically thick accretion disk, the variability is reduced and can be of the order of 1\% (see Fig. \ref{fig:hardsoft}).

\subsection{Quasi-Periodic Oscillations: low frequencies}

Quasi-periodic Oscillations (QPO) peaked components in the power-density spectra, offer a more direct way to measure frequencies, as their centroid is unambiguous, unlike the case of broad noise components (see \cite{bpk}). Since the first discovery in 1985, low-frequency (0.1-20 Hz) QPO have been detected in both classes of sources and classified in a complex way (see \cite{casellaABC}). There are strong indications that here too we are observing the same phenomenon across classes of sources. Their typical quality factor Q (defined as the ration between the frequency and the width of a PDS peak) is around $\sim$10, their fractional rms increases with energy and can be as high as $\sim$20\% above 10 keV.
In black-hole binaries, one of these QPOs (type-B) is found not to be associated to the presence of a strong noise component, which is consistent with being the evolution to higher fluxes of those shown before, and appears to be connected to the intervals of ejection of relativistic jets from the system (see \cite{fhb,belloni2010}). During the evolution of a transient source, this QPO appears in a very specific high-flux state and is of transient nature itself. Whether and how this feature is associated directly to the jet formation or ejection we do not know, but the association with the time of jet ejection, although not exact, indicates that major changes take place in the accretion flow around that time. This is also confirmed by spectral analysis (\cite{sara}).

\begin{figure} 
\begin{center}
\includegraphics[width=.98\textwidth]{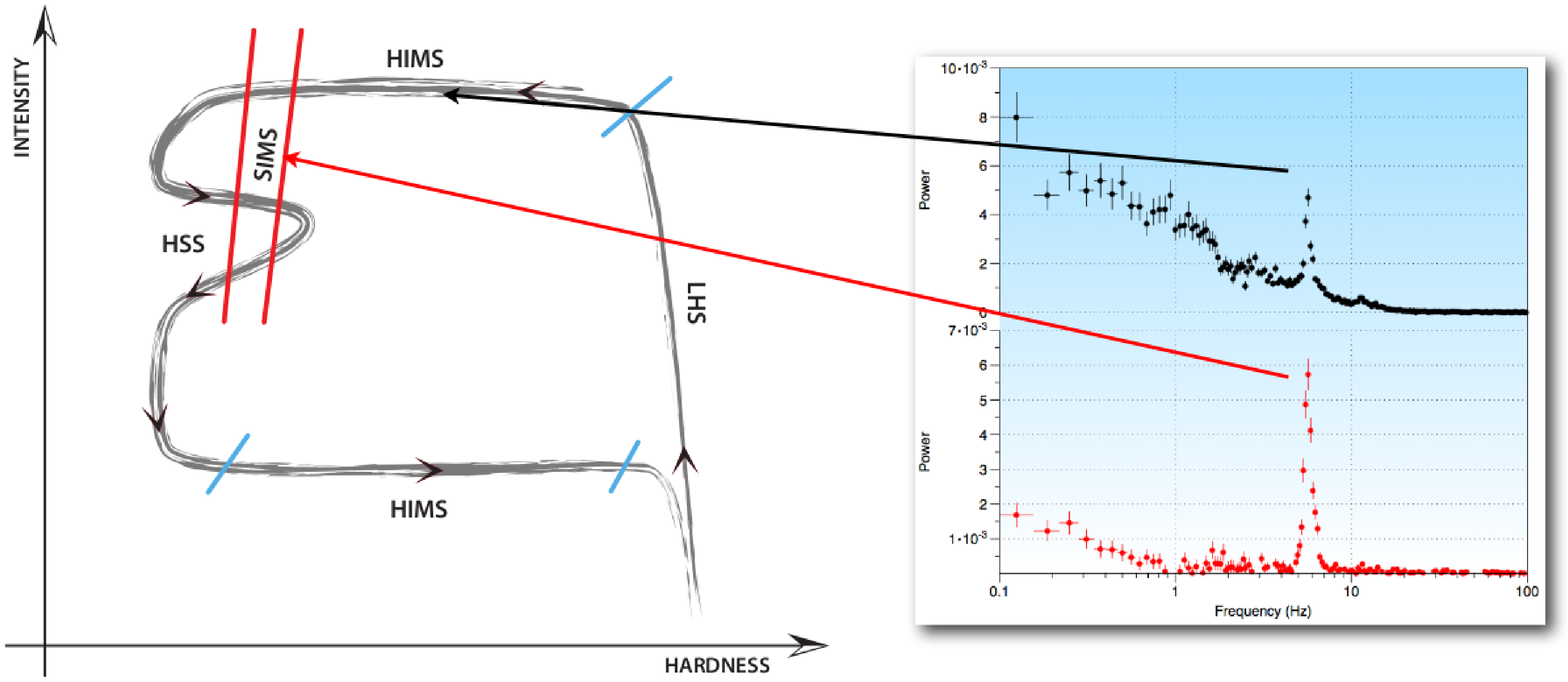} 
\caption{
Left panel: sketch of a hardness-intensity diagram of an outburst of an X-ray transient (the equivalent of a color-magnitude diagram) with the four source states marked (see \cite{belloni2010}). Right panel: two types of QPO (intentionally selected to have the same frequency) corresponding to two  high-flux states. At the top type-C QPO, associated to strong noise, at the botton type-B QPO, with reduced noise. The HIMS-SIMS transition is marked by the sudden change between these two QPOs.
}
\label{fig:hid} 
\end{center}
\end{figure}

The most common type (called horizontal-branch QPO for NS and type-C QPO for BH), has a frequency that correlates extremely well with those of the broad-band noise components, which suggests a common origin (see \cite{wk}). These low-frequency features have received less attention than high-frequency QPOs in the past decade, but are strongly connected to them and offer an insight on the accretion phenomenon, while being possibly associated to fundamental relativistic frequencies.

\subsection{Kilohertz QPOs}

The major fast time variability features are without doubt kilohertz QPOs in neutron-star LMXB. They are quasi-periodic signals in the range 200-1200 Hz. Although their full phenomenology is rather complex (see \cite{michiel2006} and references therein), we can identify the following basic features:

\begin{figure} 
\begin{center}
\begin{tabular}{cc}
\includegraphics[width=.45\textwidth]{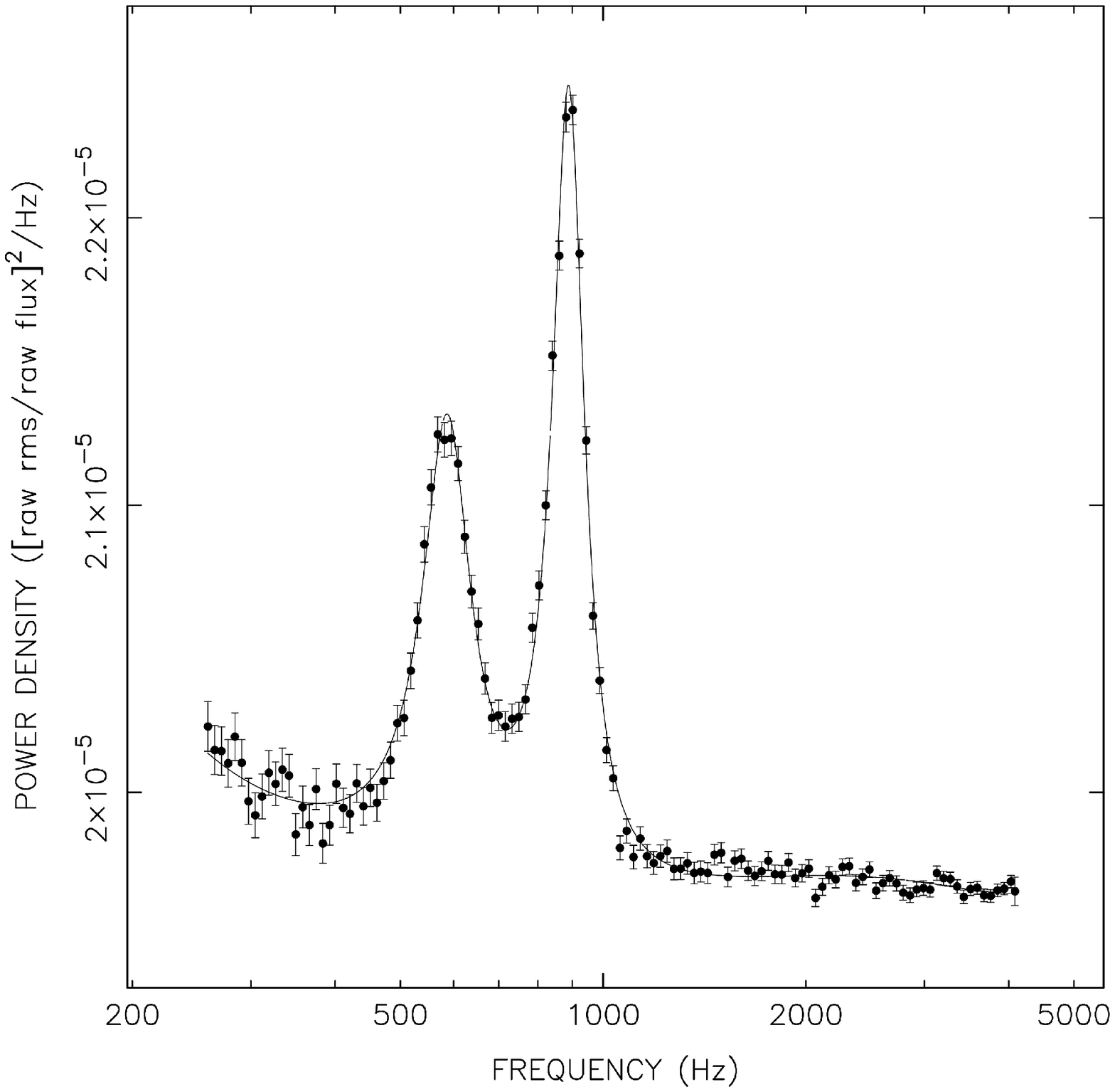} & \includegraphics[width=.5\textwidth]{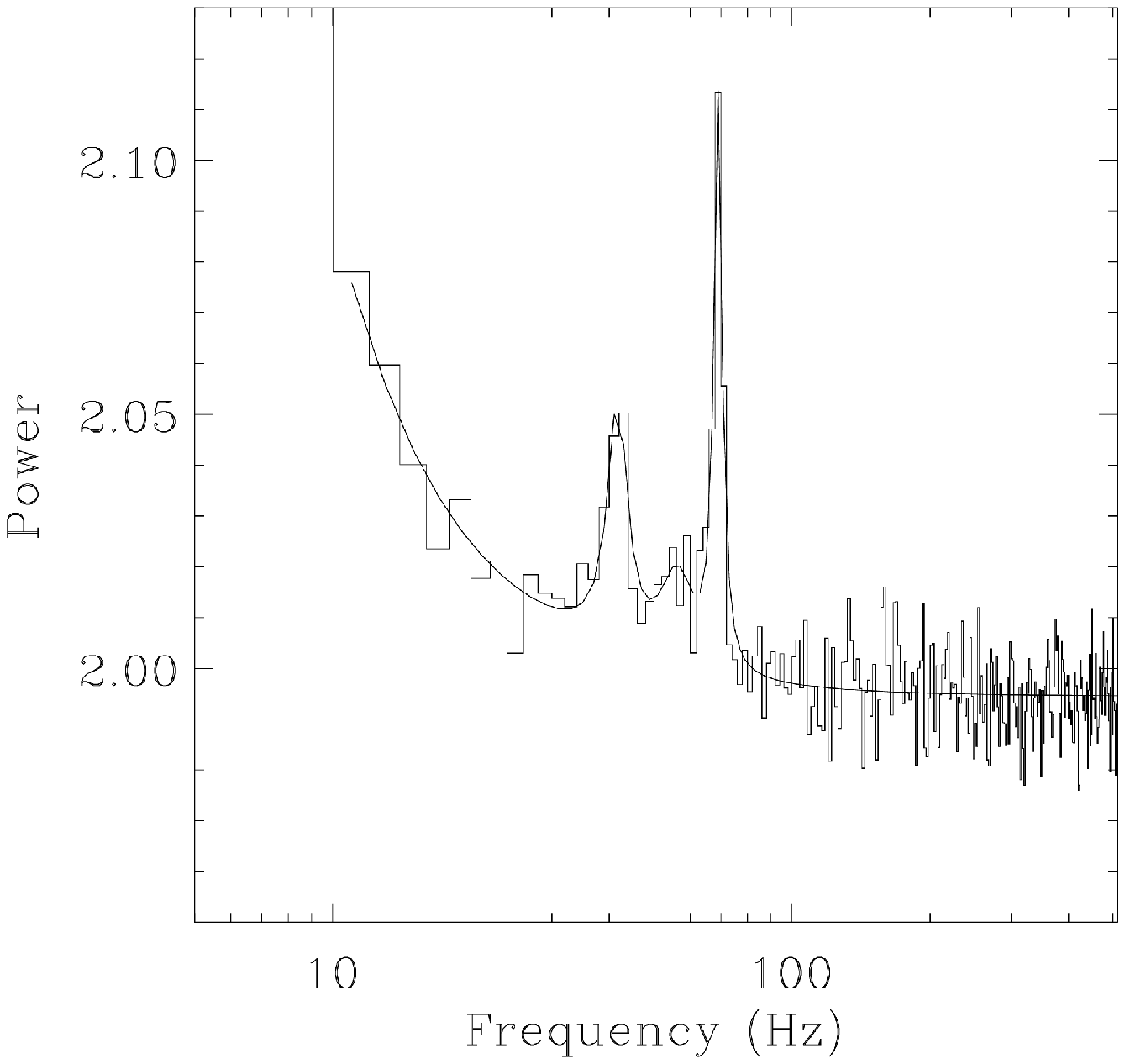}\\
\end{tabular}
\caption{
Left panel: kHz QPOs observed from the brightest LMXB Sco X-1 (from \cite{michielsco}). Right panel: two high-frequency QPOs from the black-hole binary GRS~1915+105, with centroid frequencies in 3:4 ratio (from \cite{tod1915}).
} 
\label{fig:peaks} 
\end{center}
\end{figure}

\begin{figure} 
\begin{center}
\includegraphics[width=.98\textwidth]{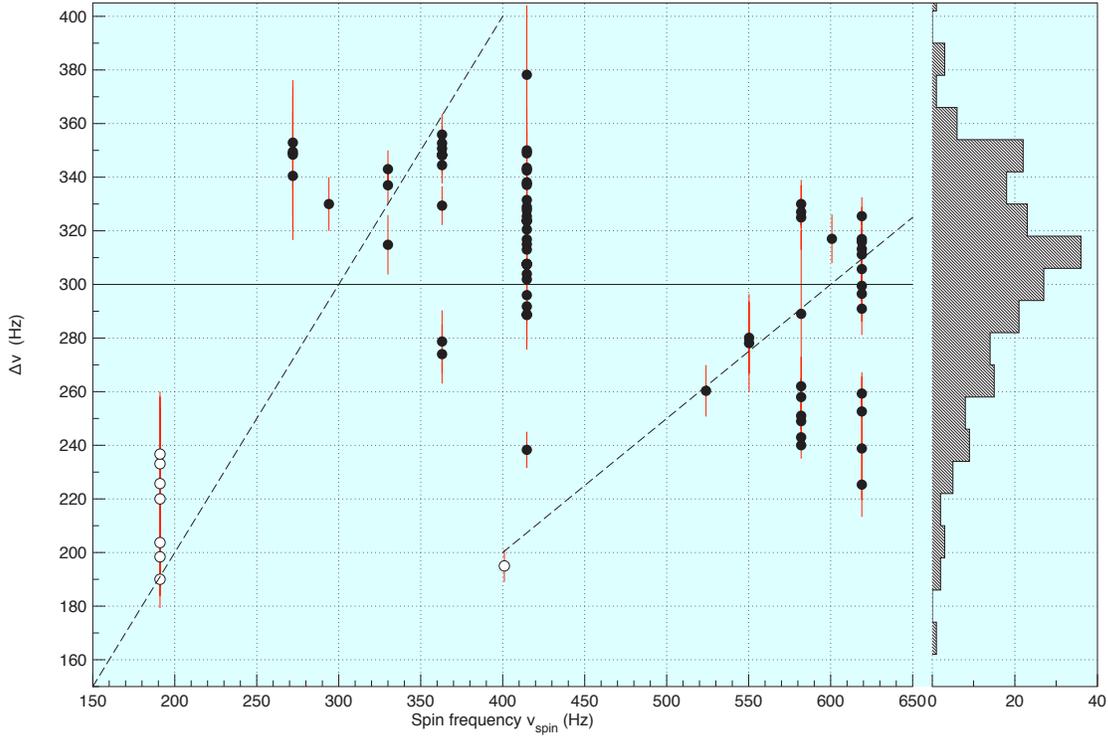} 
\caption{
Left panel: separation of the two kHz QPO frequencies $\Delta\nu$ versus the NS spin frequency for all published measurements. The white circles represent two millisecond X-ray pulsars. The solid line is a constant at $y=300$Hz, the dashed line is the ``step'' function described in the text.
Right panel: distribution histogram of all published $\Delta\nu$ values, which are more than the points in the left panel since there are sources for which the spin frequency is not (yet) known.
}
\label{fig:deltanu} 
\end{center}
\end{figure}

\begin{itemize}

\item Often, two separate peaks are observed simultaneously (see Fig. \ref{fig:peaks}), usually with different coherence. The Q value can reach values as high as 200.

\item Both peaks change frequency as a function of time in a random-walk fashion, i.e. without jumps. There are no special frequencies in the 200-1200 Hz range (\cite{bmh,saraeva,didier2}).

\item Their frequency correlates positively with source luminosity, but there is no one-to-one overall correlation. In other words, during each observation this correlation is observed, but the frequencies always cover the same range (\cite{parallel}).

\item The frequency separation between the two peaks is not constant, but for single sources is observed to decrease at high and low QPO frequencies. However, the distribution of observed separations is a rough Gaussian with average 300 Hz (see right panel in Fig. \ref{fig:deltanu} \cite{deltanu}).

\item A correlation between frequency separation and pulse frequency has been suggested, in which $\Delta\nu$ is equal to $\nu_{spin}$ if$\nu_{spin} < 400$ Hz and to half of it if $\nu_{spin} >400$ Hz. This is in contrast to the previous statement of a Gaussian distribution in $\Delta\nu$ values across all sources, which suggests that there is no connection with $\nu_{spin}$. In the left panel of Fig. \ref{fig:deltanu} one can see the plot of $\Delta\nu$ vs. $\nu_{spin}$ for all published values, to be compared with a constant value (solid line) and the ``step'' model described above. Notice that for each source, and hence for each spin frequency, there can be a number of  measurements of $\Delta\nu$ which are statistically inconsistent with each other, so that neither of these lines can be considered a fit to the data.

\item kHz QPO correlate linearly with low-frequency QPOs (of the type-C/horizontal branch flavor). The correlation between the first kHz QPO (lower) and the low-frequency QPO has been extended to low-luminosity systems and black-hole binaries, provided noise frequencies are used for those (\cite{pbk,bpk}).

\end{itemize}

Given the high frequencies, it is clear that these signals come from the innermost regions of the space-time around the neutron star. Although the basic phenomenology outlined above is rather clear, theoretical models are still not able to reproduce it.

One important point is whether we can use these signals to find direct evidence of the presence of an innermost stable circular orbit (ISCO) around the neutron star. If the highest-frequency QPO  observed in a system is associated to a Keplerian frequency at the ISCO, we can use it to measure the ISCO itself. Recently, effects related to the ISCO have been claimed, although there is still debate on possible alternatives (\cite{didier,mendez}).

Since another approach to this measurement is through the analysis of relativistically broadened iron fluorescence lines, it is interesting to compare the results of the two methods on simultaneous data. Obtaining such data is not simple, but (quasi)-simultaneous RXTE/Chandra/XMM data exist for the source 4U~1636--53. From these data, it appears that the two measurements can be reconciled only with a massive neutron star ($> 2 M_\odot$ or $> 3 M_\odot$ depending on the line model used), indicating that more observational data are needed in order to obtain a robust comparison (see \cite{diego1636}).

\subsection{Black-hole high-frequency QPOs}

For black-hole binaries, the situation at high frequencies (above 30 Hz) is very different. Despite the extremely large base of data covering outbursts of transient sources and monitoring of persistent sources, only very few detection of high-frequency features have been obtained (see \cite{bellonisoleri}). In all seven sources where narrow peaks were found (and only in few observations), the frequency seems to be constant. Moreover, in four sources two peaks have been detected simultaneously, although not always in the same energy band. In three of these cases their frequencies are in ratio 2:3, in the fourth in ratio 3:4 (see Fig. \ref{fig:peaks}). There is an indication that the highest of the two QPO frequency is anti-correlated with the mass of the black hole, as expected if it was a Keplerian frequency at the ISCO.
Thee features are observed in or close to a particular source state, the same one when type-B LFQPO are observed. However, even within this state only a few observations yield a detection.
There seems to be no correlation with the frequency of low-frequency features, although a comparative statistical study has not been done yet.

\section{Theoretical models}

All the observables described above need to find a physical explanation. QPOs provide very precise frequencies to compare with models, but it is becoming clear now that other variables, such as the quality factor and the fractional rms of the peaks must be considered in order to understand the origin of the oscillations (see \cite{didier,mendez}).

The original class of models for QPOs was based on the beat between the neutron star rotation frequency and the Keplerian frequency of accreting matter at a special radius (see \cite{mlp,michiel2006}). About a decade ago, the relativistic frequencies were introduced by the Relativistic Precession Model (cite{stellavietri}). The LFQPO (of type-C) and two kHz QPO are identified with Lense-Thirring precession, nodal precession and Keplerian motion respectively. In the model, only the basic identification with fundamental relativistic frequencies is made, but the comparison with observations is rather successful and indicates that this direction is very promising for a more thorough understanding. 

For high-frequency QPOs in black-hole binaries, the relativistic resonance models interpret the fixed frequencies in terms of resonance between orbital and epicyclic frequencies at a particular radius, yielding special ratios between the frequencies (see \cite{michiel2006} for a discussion). More detections are needed to confirm this model, at variance with the case of neutron stars, where more models are needed to explain the large set of observational data.

In addition to probing accretion and General Relativity, kHz QPOs can be used to put useful limits on the equation of state of neutron matter (see \cite{miller1999}). The fact that matter is orbiting with period $P$ at a radius $R$ around the neutron star puts a limit to the relation between mass and radius of the star, which should be smaller than $R$. In addition, $R$ cannot be smaller than the ISCO. With the highest QPO frequency, limits can already be put, but the observability of a higher frequency would make these limits very stringent. Unfortunately, although the instrumental sensitivity would allow to detect signals well above $\sim$1.2 kHz, the rms and coherence of kHz QPOs decrease fast (\cite{didier}) and no detection is available above that frequency.

\section{The future}

\begin{figure} 
\begin{center}
\includegraphics[width=.98\textwidth]{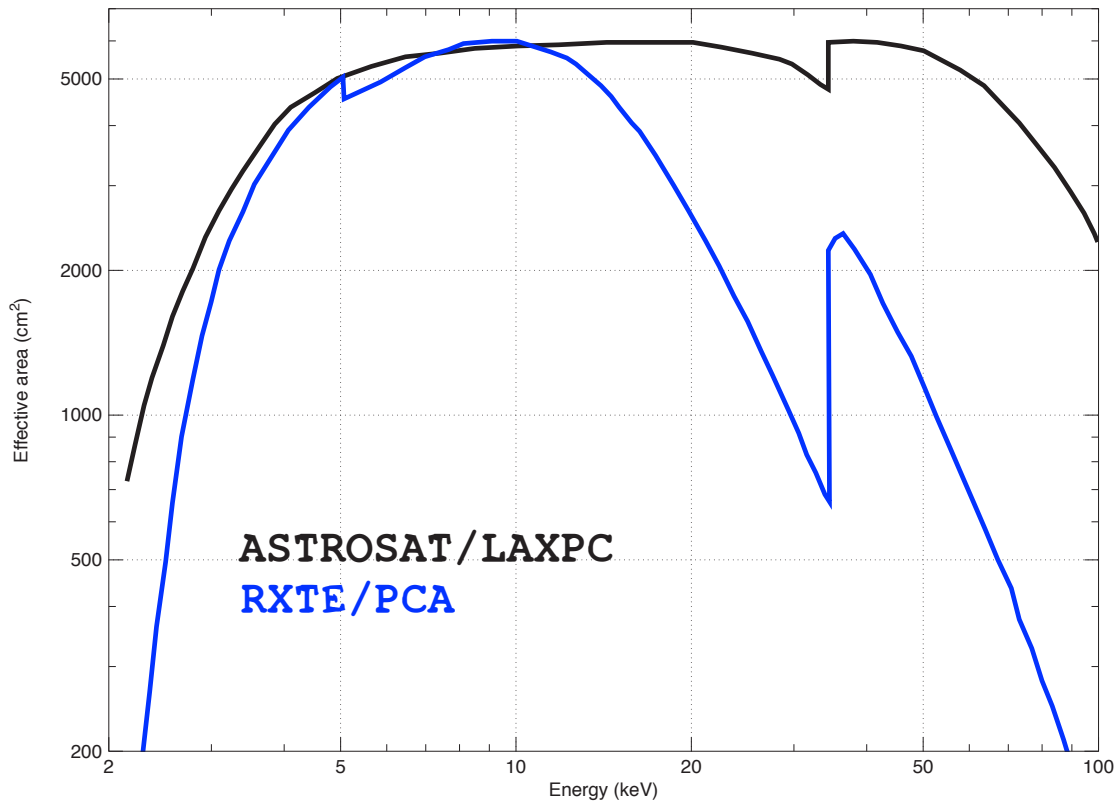} 
\caption{
Comparison between the effective areas of the ASTROSAT/LAXPC (black) and the RXTE/PCA (five PCU elements, blue). 
}
\label{fig:laxpc} 
\end{center}
\end{figure}

The space mission that brought all advancements in fast timing for X-ray binaries is the Rossi X-Ray Timing Explorer. After 15 years of service, the satellite will soon cease operations. There are a number of proposed instruments and missions devoted to fast timing, such as AXTAR and the HTRS on board the International X-ray Observatory, all on medium to long time scales. Other papers in these proceedings deal with them.
However, in the spring of 2011 the indian satellite for astronomy ASTROSAT will be launched\footnote{\tt http://meghnad.iucaa.ernet.in/\textasciitilde astrosat/}. Among other instruments covering a wide energy range from the ultraviolet to hard X-rays, ASTROSAT will carry a large high-pressure proportional counter (LAXPC) which will provide a response similar to that of the RXTE/PCA below 10 keV and much higher at higher energies (see Fig. \ref{fig:laxpc}). This constitutes a unique possibility to continue timing studies, the major advantage that the collecting area, which even below 10 keV is better than that of the current PCA, where obtaining a 5-unit observation is very unlikely, resulting in a considerable increase in area. Moreover, most of the timing features shown before are more intense at high energies, where the increased area of the LAXPC will bring a higher sensitivity. In particular, it will be possible to study black-hole high-frequency QPOs with higher sensitivity, hopefully bringing the necessary increase in observational wealth to compare them in details with theoretical models. 

\end{document}